\documentclass[aps,pre,epsf,superscriptaddress,amsmath,amssymb,amsfonts,twocolumn,showpacs]{revtex4-1}

\usepackage{graphicx}
\usepackage{epsfig}
\usepackage{dcolumn}
\usepackage{bm}
\usepackage{braket}
\usepackage{amsmath}
\usepackage{color}
\newcommand{\abs}[1]{\left| #1 \right|}
\usepackage[utf8]{inputenc}
\usepackage[left=2cm,right=2cm,top=0.0cm,bottom=2cm,includeheadfoot]{geometry}

\usepackage{hyperref}
\usepackage{multirow}
\DeclareMathOperator{\sech}{sech} 
\usepackage{todonotes}

\begin{document}
\title{Observation and Analysis of Multiple Dark-Antidark Solitons \\ in Two-Component 
Bose-Einstein Condensates}

\author{G. C. Katsimiga}
\affiliation{Center for Optical Quantum Technologies, Department of Physics,
University of Hamburg, Luruper Chaussee 149, 22761 Hamburg,
Germany}

\author{S. I. Mistakidis}
\affiliation{Center for Optical Quantum Technologies, Department of Physics,
University of Hamburg, Luruper Chaussee 149, 22761 Hamburg,
Germany}

\author{T. M. Bersano}
\affiliation{Department of Physics and Astronomy, Washington State University, Pullman, WA 99164-2814, USA}

\author{M. K. H. Ome}
\affiliation{Department of Physics and Astronomy, Washington State University, Pullman, WA 99164-2814, USA}

\author{S. M. Mossman}
\affiliation{Department of Physics and Astronomy, Washington State University, Pullman, WA 99164-2814, USA}

\author{K. Mukherjee}
\affiliation{Center for Optical Quantum Technologies, Department of Physics,
University of Hamburg, Luruper Chaussee 149, 22761 Hamburg,
Germany}\affiliation{Indian Institute of Technology Kharagpur, Kharagpur-721302, West Bengal,
India}

\author{P. Schmelcher}
\affiliation{Center for Optical Quantum Technologies, Department of Physics,
University of Hamburg, Luruper Chaussee 149, 22761 Hamburg,
Germany} \affiliation{The Hamburg Centre for Ultrafast Imaging,
University of Hamburg, Luruper Chaussee 149, 22761 Hamburg,
Germany}

\author{P. Engels}
\affiliation{Department of Physics and Astronomy, Washington State University, Pullman, WA 99164-2814, USA}

\author{P. G. Kevrekidis}
\affiliation{Department of Mathematics and Statistics, University
of Massachusetts Amherst, Amherst, MA 01003-4515, USA}

\affiliation{Mathematical Institute, University of Oxford, OX26GG, UK}

\date{\today}

\begin{abstract}
We report on the static and dynamical properties of multiple dark-antidark
solitons (DADs) in two-component, repulsively interacting Bose-Einstein condensates.
Motivated by experimental observations involving multiple DADs, we present a theoretical study 
which showcases that bound states consisting of dark (antidark) solitons in the first (second) component 
of the mixture exist for different 
values of interspecies interactions. It is found that ensembles of few DADs may exist as stable 
configurations, while for larger DAD arrays, the relevant windows of stability with respect to the 
interspecies interaction strength become progressively narrower.
Moreover, the dynamical formation of states consisting of alternating
DADs in the two components of the mixture is monitored.
A complex dynamical evolution of these states is observed, leading either to sorted DADs or 
to beating dark-dark solitons depending on the strength of the interspecies coupling. 
This study demonstrates clear avenues for future investigations of DAD configurations.
\end{abstract}

\maketitle

\section{Introduction}

Over the past 25 years, the experimental implementation of dilute gas Bose-Einstein condensates 
(BECs) has provided a fertile platform for the
exploration of a wide range of macroscopic quantum
features~\cite{becbook1,stringari,emergent}.
One of the major axes around which this effort has revolved is the study
of nonlinear waves and their existence, dynamics and interactions
within this atomic physics
platform~\cite{siambook}.
Such waves were previously recognized as playing a substantial role in other fields including 
nonlinear optics~\cite{DSoptics}, as well as water waves~\cite{ablowitz}.
The realm of atomic BECs, however, has enabled the study of a wide range of such patterns 
including, but not limited to, dark solitons~\cite{Frantzeskakis_2010}, vortical states~\cite{Alexander2001,fetter2,komineas} and also more complex entities such as 
hopfions~\cite{Hopfion1} and even vortex knots~\cite{PhysRevE.85.036306,PhysRevA.99.063604}.

Beyond the setting of single component atomic condensates (which have been the focus of many of the above studies), recent efforts have considered multi-component generalizations of soliton formation and dynamics~\cite{Kevre}.
A fundamental state of interest has been the dark-bright (DB) solitary wave and its close relative, the dark-dark solitary state~\cite{christo,vdbysk1,vddyuri,ralak,dbysk2,shepkiv,parkshin}.
While the relevance of this state to the dynamics of multiple polarizations of light in photorefractive crystals was originally recognized early on in pioneering works in nonlinear optics~\cite{seg1,seg2}, it was not until their proposal~\cite{buschanglin} and especially first experimental realization~\cite{hamburg} in atomic BECs that an explosion of interest ensued~\cite{pe1,pe2,pe3,pe4,pe5,azu}.
In recent years, this direction of research has gained further momentum through the study of variants of solitary waves such as nonlinear polarization waves~\cite{kamch} and magnetic solitons~\cite{qupit}, so-called dark-antidark structures (where antidark means a bright solitonic state on top of a finite background)~\cite{Danaila}, and extensions to spinor (three-component) solitary waves~\cite{uspeter}.
This thrust continues intensely through both experimental and theoretical studies; see, e.g., Ref.~\cite{rueda} for a recent synthesis.

In an earlier work, we reported the possibility
of an experimental realization of dark-antidark (DAD) structures~\cite{Danaila}, as indicated above.
The experimental efforts described in the present work showcase the
generation of multiple such structures in two different formats: we have observed settings in which the (multiple) dark solitons are all in the same component and the antidark solitons are all in the second component, as well as ones where there is an alternating sequence of darks and antidarks (in a complementary fashion) between the two components.
This, in turn, motivates a theoretical study of each one of these two configurations of multi-dark-antidark solitons.
On the one hand, this investigation extends naturally the setting of a single dark-antidark soliton~\cite{Danaila}, while on the other hand, it complements studies of multiple dark-bright solitons~\cite{pe3,wang} and even multiple dark solitons~\cite{siambook}.
The key result is that states in which all dark solitons are contained in one
component and antidarks in the other can be dynamically robust for
a few DADs but become progressively less stable for more DADs, in
line with experimental observations.
Also in agreement with experiment, alternating DAD states (where, for instance, in each component a dark soliton is neighboring two antidark solitons and vice-versa) are not found as stationary solutions, but only as dynamical states in the two-component systems considered.
The dynamics of both types of states is explored.

Our presentation is structured as follows. In Section~\ref{sec:experiment} we present
the experimental demonstration of the corresponding states as they
are generated in our two-component experiments with $^{87}$Rb BECs.
Section~\ref{theory} describes the modeling platform of the relevant system, while in Section~\ref{sorted} we analyze the existence, stability and dynamics of the different multi-soliton DAD states obtained in the context of the theoretical model.
In Section~\ref{alternating} we briefly discuss the dynamical evolution of configurations comprised of few alternating DADs.
{Finally, we summarize our findings and present our conclusions as well as some directions for future studies in Section~\ref{conclusions}.}

\section{Experimental demonstration}\label{sec:experiment}
To motivate our investigation of multi-soliton DAD states, we begin by showcasing experimental results demonstrating the creation and stability of such structures.
To experimentally realize an arrangement of DAD solitons, we employ a procedure based on Rabi winding~\cite{Matthews,Williams}.
An elongated BEC of approximately $4.5\times10^5$ $^{87}$Rb atoms is prepared in the $\lvert F,m_F\rangle=\lvert 1,-1\rangle$ hyperfine state, optically confined by a single beam dipole trap with trap frequencies of approximately $\{\omega_x,\omega_y,\omega_z\}=2\pi\{1.5, 140, 180\}$ Hz. The weakly confined $x$ axis is oriented horizontally.

\begin{figure*}
 \includegraphics{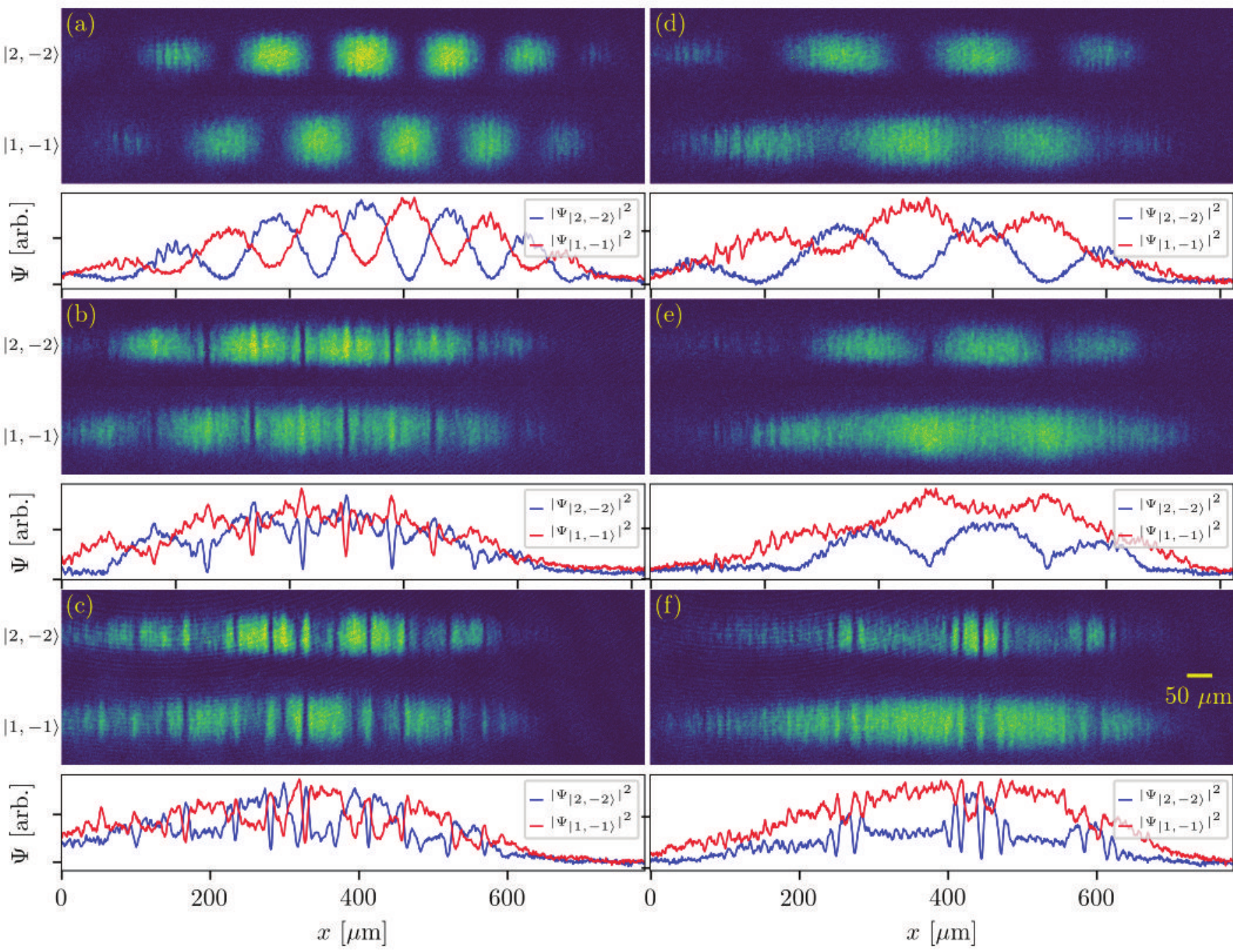}
    \caption{Absorption images of the elongated Bose-Einstein condensate after Rabi winding with (a)-(c) near zero detuning and
    (d)-(f) a detuning of $\delta=2\pi\cdot11$ kHz detuning. A dual imaging procedure places the $\lvert2,-2\rangle$ atoms above the 
    $\lvert1,-1\rangle$ atoms for each measurement. 
    The images (a), (d) are taken after 15 ms of MW driving, (b), (c) after an additional 150 ms of
    undriven evolution, and (c), (f) after 410 ms. 
    The corresponding integrated cross sections of each of the two components are shown for each panel.}
    \label{fig:BEC}
\end{figure*}

We apply an external magnetic field of approximately 1 G with a linear gradient of approximately 5 mG/cm along the long axis of the BEC. This produces a spatially varying Zeeman shift within the $F=1$ and $F=2$ hyperfine manifolds. A uniform, fixed frequency microwave (MW) field is used to drive transitions between the $\lvert 1,-1\rangle$ and $\lvert 2,-2\rangle$ states. In the following, we consider these two states as the pseudo-spin orientation of a spin-1/2 system. The two-level Rabi formula for the population of atoms in the
$\lvert 2,-2\rangle$ level, $P_{\lvert 2,-2\rangle}$, takes a spatially dependent form~\cite{Deb} given by
\begin{equation}
    P_{\lvert 2,-2\rangle}(x,t)= \frac{\Omega^2}{\Omega^2+\delta^2(x)} \sin^2\left(\sqrt{\Omega^2+\delta^2(x)}t/2\right),
    \label{eq:rabi}
\end{equation}
where $\Omega$ is the resonant Rabi frequency, $t$ is the winding time, and $\delta$ is the detuning of the microwave coupling which varies across the cloud due to the spatially dependent Zeeman shift.
By applying the microwave drive for a given amount of time, a magnetization pattern is created in the
condensate corresponding to the phase winding caused by the differential Rabi cycling.
The magnetic field gradient is then suddenly removed and the BEC is allowed to evolve in the now homogeneous magnetic field.
The resulting dynamics are observed by turning off the optical trap and selectively imaging the two hyperfine states sequentially after 15 ms (17 ms) of time of flight for the $\lvert 2,-2\rangle$ ($\lvert 1,-1\rangle$) state. Our imaging technique allows us to image both states in each experimental run.
There is negligible evolution occurring in the time between the two
exposures taken for each run, so that features in the two states can
be accurately superimposed to represent the in-trap spin structure of
the state. Fig.~\ref{fig:BEC} presents a variety of these absorption
images where the $\lvert 2,-2\rangle$ state is on top and the $\lvert
1,-1\rangle$ is at the bottom, followed by integrated cross sections corresponding to the absorption image above.

For a microwave driving frequency that is close to the resonance frequency at the center of the atom cloud, a high amplitude 
winding pattern results.
Fig.~\ref{fig:BEC}(a) shows a state produced after 15 ms of Rabi winding with $\Omega/2\pi=19$ kHz. 
The population of the two (pseudo-) spins alternates along the elongated direction of the BEC with a large amplitude. 
Looking at Fig. 1, it is important to reiterate that the two spin components are imaged 
separately while the cloud falls, but the dynamics we describe here occur in the optical trap 
where the two spin components exist together. 
The bright features in one spin component fill the dark regions of the other spin component in 
this case. 
After 150 ms of evolution in the optical trap, which is a long time on the scale of 
mean field effects but a short time in terms of the $x$-axis trap frequency, a regular array of 
alternating DAD solitons corresponding to the initial winding appear (see Fig.~\ref{fig:BEC}(b)). 
The DAD solitons are characterized by a dark notch of low density in one component, which is 
filled in by a bright stripe of high density in the other component. 
These features are distinct from dark-bright solitons in that the bright component exists on a 
finite background density which does not go to zero. 
After allowing these solitons to evolve in the trap for 450 ms, which is longer than 
half of a trap period, Fig.~\ref{fig:BEC}(c) shows that the arrangement has broken into an irregular
collection of solitonic features and domains rather than a regular array of alternating DAD solitons. 
One can clearly discern some well-defined features with widths characteristic of dark solitons~\cite{Hamner}, 
but the regular pattern demonstrated in Fig~\ref{fig:BEC}(b) does not persist. 

Alternatively, a MW driving frequency which is farther detuned from resonance will produce an 
array of spin mixed regions separated by regions of spin purity, as described by Eq.~(1) when the 
detuning $\delta$ varies linearly with a large offset. Fig.~1(d)-(f) show the spin populations 
after an identical winding procedure to that describing Fig.~1(a)-(c), and for same time 
increments, but with a microwave frequency 11 kHz detuned from the resonant 
frequency at the center of the cloud. Here, the partial spin transfer only allows dark solitons to 
form in the $\lvert 2,-2\rangle$ component. After approximately 250 ms of evolution time following the 
microwave winding, collections of DADs start to nucleate leading to persistent configurations 
containing between two and four DAD solitons in clusters as shown in Fig.~1(f). 
Fig.~1(f) shows density notches in the $\lvert 2,-2\rangle$ component with the width characteristic of dark solitons in this 
system and there are corresponding high density regions in the other spin state. While we observe 
density depressions in the second component ($\lvert 1,-1\rangle$) between the antidark solitons, 
these depressions are of a larger length scale than the characteristic soliton scale and are 
therefore more appropriately interpreted as a reduced background density for the antidark 
solitons. These clusters are unique to the non-alternating (alias sorted) DAD configurations, i.e. those with all the 
dark features in one component and bright features in the other, and therefore motivate a detailed 
investigation of the conditions under which stable DAD configurations can form.

\section{Theoretical Framework}\label{theory}

Motivated by the experimental observations of Section~\ref{sec:experiment}, we now proceed to describe the theoretical framework for 
modelling these solitonic configurations.
We consider a binary mixture of repulsively interacting BECs
composed of the two hyperfine states mentioned in the previous section, namely $\ket{F=1, m_F=-1}$ and $\ket{F=2, m_F=-2}$,
of $^{87}$Rb~\cite{Egorov} being
confined in a one-dimensional (1D) harmonic oscillator potential.
Such a cigar-shaped geometry can be realized
experimentally~\cite{Becker,Hoefer,Middelkamp}
in a highly anisotropic trap with the longitudinal
and transverse trapping frequencies obeying
$\omega_x \ll \omega_{\perp}$, as described in  Section~\ref{sec:experiment}.
Within mean-field theory the dynamics of this binary mixture can be
well approximated by the following system of coupled Gross-Pitaevskii equations (GPE) 
of motion~\cite{stringari,emergent,siambook}:
\begin{eqnarray}
i\hbar \partial_t \psi_j =
\left( -\frac{\hbar^2}{2m} \partial_{x}^2 +V(x) -\mu_j + \sum_{k=1}^2
g_{jk} |\psi_k|^2\right)\!\psi_j.\nonumber\\
\label{model}
\end{eqnarray}
In the above expression, $\psi_j(x,t)$ ($j=A,B$) denotes the wavefunction for 
the $A\equiv\ket{1, -1}$ and $B\equiv\ket{2, -2}$ hyperfine states respectively, and $g_{jk}$ is 
the interaction coefficient between species $j$ and $k$.
Note here that in this framework we do not consider particle transfer between the components
of the mixture since they lie in a different spin manifold and therefore such processes
are negligible as it has also been confirmed experimentally.
Each $\psi_j(x,t)$ is normalized to the
corresponding number of atoms i.e. $N_j = \int_{-\infty}^{+\infty} |\psi_j|^2 dx$.
Also, $m_A=m_B=m$ and $\mu_j$ refer to the atomic mass and chemical potentials for each of the
species respectively.
The effective 1D coupling constants are given by $g_{jk}=2\hbar\omega_{\perp} a_{jk}$,
where $a_{jk}$ are the three $s$-wave scattering lengths
(with $a_{AB}=a_{BA}$)
accounting for collisions between atoms that belong to
the same ($a_{jj}$) or different ($a_{jk}, j \ne k$) species.
We note that both the intra- and interspecies scattering lengths can, in principle, be manipulated 
experimentally by means of Feshbach~\cite{kohler,Chin}
or confinement induced resonances~\cite{Olshanii,Kim}.
Finally, $V(x)$ represents the external trapping potential.

For the numerical analysis presented below, we express the system of Eqs.~(\ref{model})
in the following dimensionless form:
\begin{align}\label{eqn::GPE_A}
i \frac{\partial \psi_A}{\partial t} &= \left[ -\frac{1}{2} \frac{\partial^2}{\partial x^2} + 
V+ g_{AA} |\psi_A|^2 + g_{AB} |\psi_B|^2 - \mu_A\right] \psi_A\\\label{eqn::GPE_B}
i \frac{\partial \psi_B}{\partial t} &= \left[ -\frac{1}{2} \frac{\partial^2}{\partial x^2} + V
+ g_{BB} |\psi_B|^2 + g_{AB} |\psi_A|^2 - \mu_B\right] \psi_B.
\end{align}
Here $\mu_i$ is the chemical potential of the $i$-th species and
$V(x)=\frac{1}{2} \omega_{1}^2 x^2$ is the dimensionless harmonic trapping potential with
$\omega_A=\omega_B \equiv\omega_{1} = \omega_x/\omega_\perp=0.1$.
Similar results can be found for much more elongated condensates with
trap ratios $\omega_1=0.01$ closer to the one of the above
reported experiments.
For the numerical findings to be presented below,
we fix the inter- and intraspecies interaction coefficients
to the experimentally relevant values of the hyperfine
states of $^{87}$Rb, namely, $g_{AA}=1.004$ and $g_{BB}=0.9898$.
To show the formation and variation of the
dark-antidark structures of interest, we first examine the relevant solutions as a function of $g_{AB}$, and then focus on
parameter values proximal to the experimental ones. 
In the dimensionless units adopted above densities $|\psi_i|^2$, length, energy and time are measured in units of
$2a_{AA}$, $\sqrt{\frac{\hbar}{m \omega_\perp}}$, $\hbar \omega_\perp$ and $\omega_\perp^{-1}$ respectively.

\begin{figure}[htb]
 \centering \includegraphics[width=0.47\textwidth]{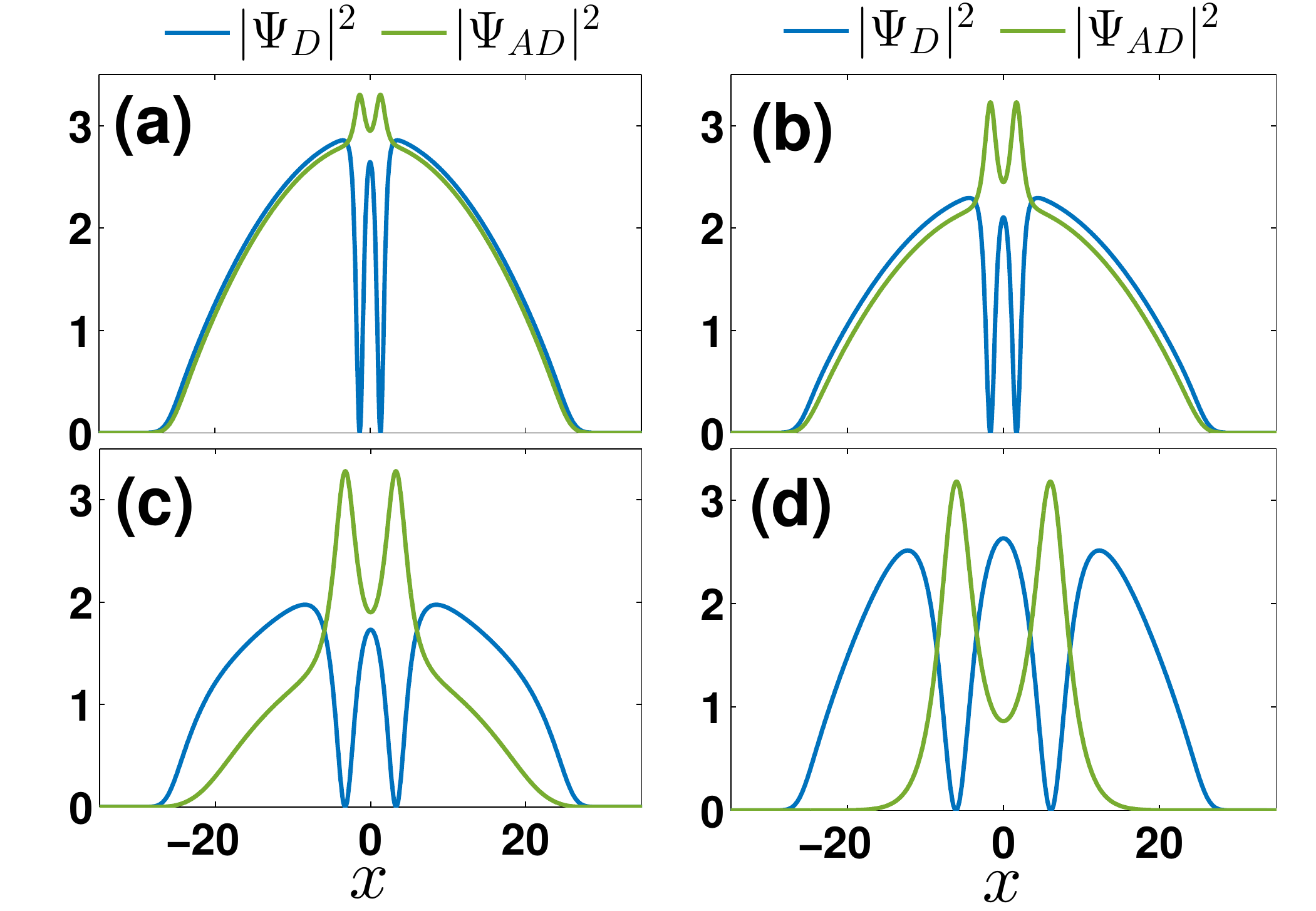}
 \caption{\label{Figure_2DADs_stat} Stationary states of
 two sorted DAD configurations for interspecies couplings (a) $g_{AB}=0.2$, (b) $g_{AB}=0.5$,
 (c) $g_{AB}=0.9$ and (d) $g_{AB}=0.98$.
 The densities $|\Psi_D|^2$ of the dark and $|\Psi_{AD}|^2$ of the antidark solitons are
 illustrated (see legend).
 In all cases the states remain stable for long evolution times up to $t=10^4$.
 Other parameters correspond to $\omega_1=0.1$, $g_{AA}=1.004$, $g_{BB}=0.9898$, $\mu_A=3.5$
 and $\mu_B=3.4$.}
 \end{figure}

\begin{figure*}[htb]
 \centering \includegraphics[width=0.9\textwidth]{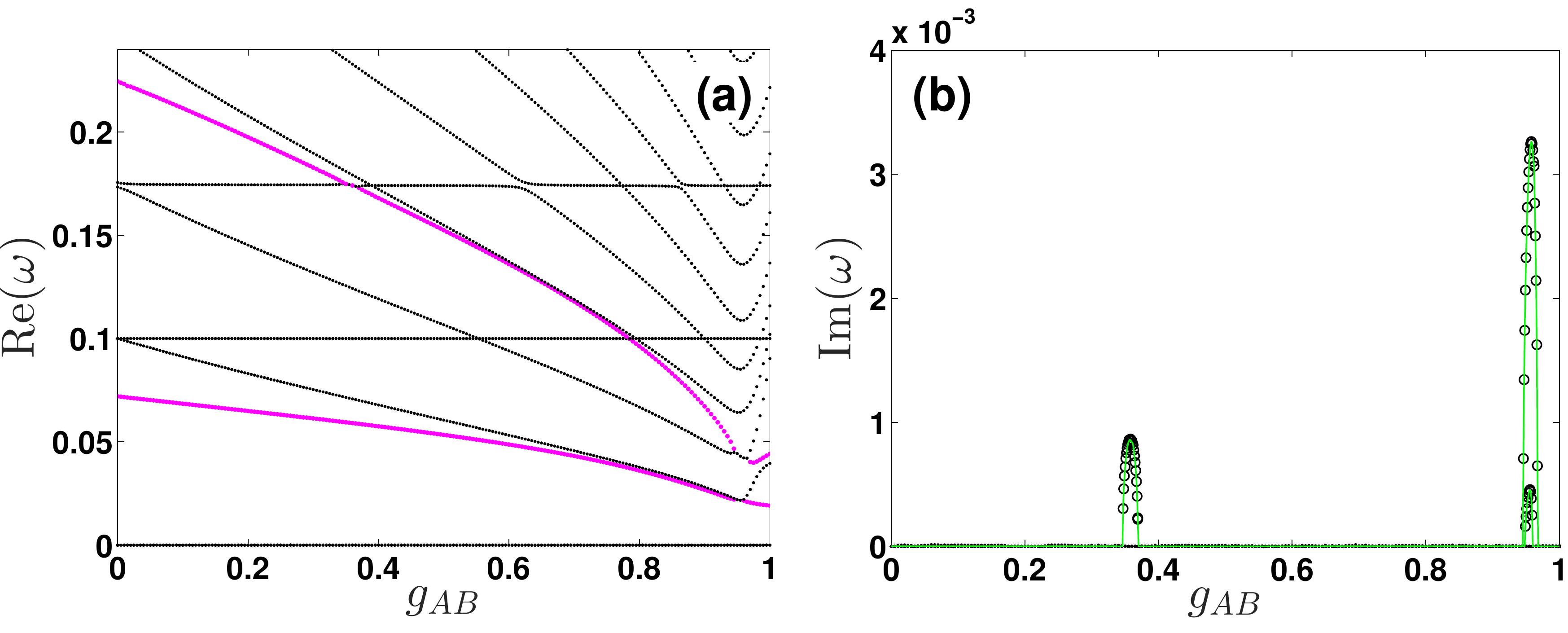}
 \caption{BdG spectrum of a stationary state composed of two
 sorted DAD solitons as a function of the interspecies interaction strength $g_{AB}$.
 (a) The real part, ${\rm Re}(\omega)$ and (b) the imaginary part, ${\rm Im}(\omega)$, of the
 underlying eigenfrequencies is shown as a function of $g_{AB}$. 
 Note that each point of the line ${\rm Re}(\omega)=0$ is quadruple degenerate due to particle  
 number conservation per component, while the existence of a finite imaginary part signals the  
 presence of an instability of the two-DAD configuration that occurs 
 for $g_{AB} \in [0.347, 0.369]$ and $[0.946, 0.967]$. 
 The solid lines in panel (b) provide a guide to the eye for each of the bifurcation loops 
 appearing in the imaginary part of the eigenfrequencies.
 Other system parameters correspond to $\omega_1=0.1$, $\mu_A=3.5$, $\mu_B=3.4$, $g_{AA}=1.004$ 
 and $g_{BB}=0.9898$. 
 \label{Figure_2DADs_BdG_g12} }
 \end{figure*}

 \begin{figure}[htb]
 \centering \includegraphics[width=0.47\textwidth]{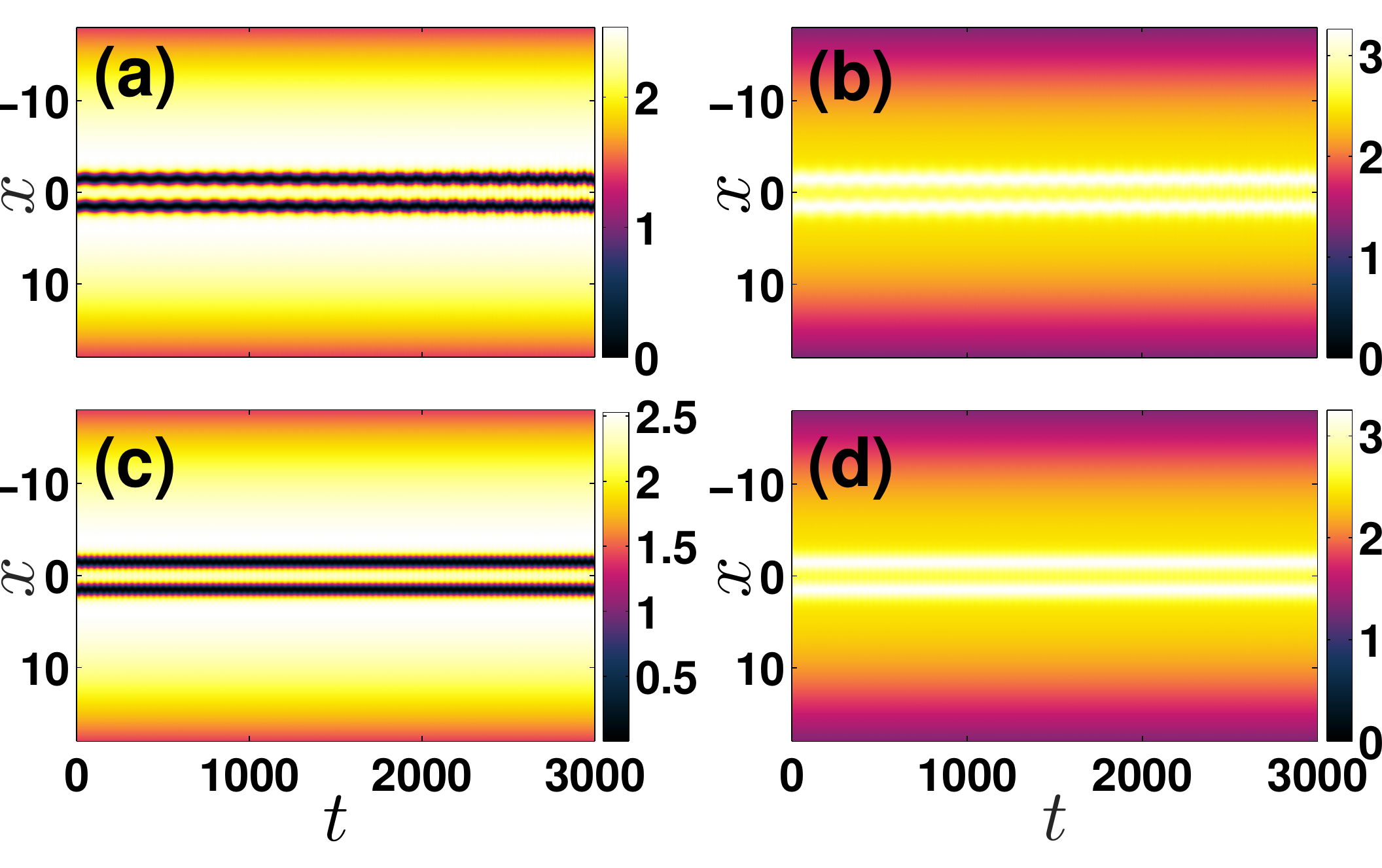}
 \caption{\label{Figure_2DADs_036} (a)-(d) Dynamical evolution of a stationary state of two sorted
 DAD solitons.
 Panels (a), (c) [(b), (d)] illustrate the evolution of the dark [antidark] soliton
 component upon adding the eigenvectors of the first [(a), (b)] and the second
 [(c), (d)] anomalous mode identified in the BdG spectrum of Fig. \ref{Figure_2DADs_BdG_g12}.
 Other system parameters correspond to $\mu_A=3.5$, $\mu_B=3.4$, $g_{AA}=1.004$, $g_{BB}=0.9898$ 
 and $g_{AB}=0.36$.}
 \end{figure}

 \begin{figure}[htb]
 \centering \includegraphics[width=0.47\textwidth]{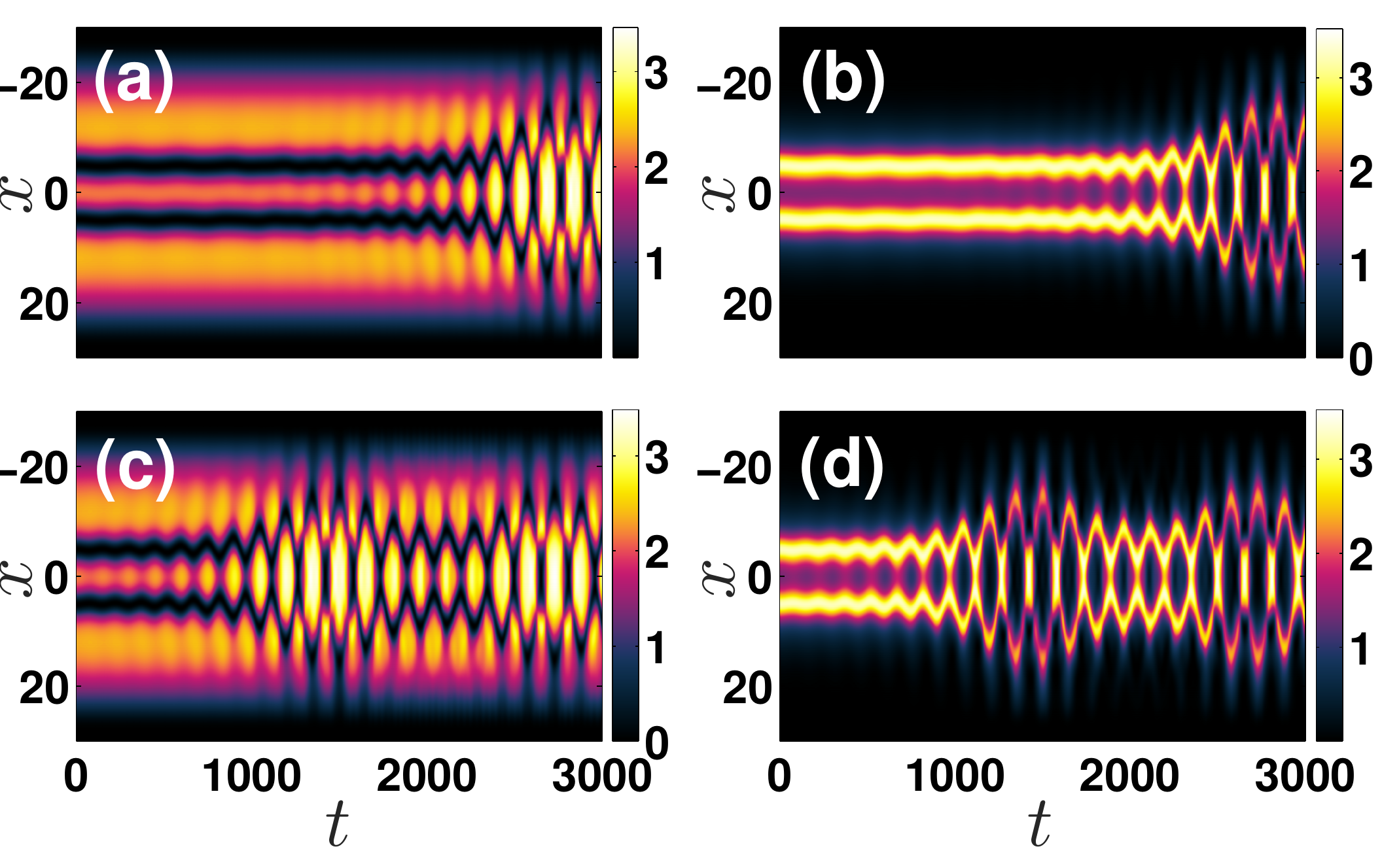}
 \caption{\label{Figure_2DADs_096}
 (a)-(d) Evolution of a stationary state of two sorted DAD solitons for $\mu_A=3.5$.
 The time-evolution of the dark [antidark] soliton components is presented 
 in panels (a), (c) [(b), (d)] upon adding the eigenvectors of the first [(a), (b)], and the 
 second [(c), (d)] anomalous mode identified in the BdG spectrum of 
 Fig.~\ref{Figure_2DADs_BdG_g12}.
 The remaining system parameters are $\mu_B=3.4$, $g_{AA}=1.004$, $g_{BB}=0.9898$ and 
 $g_{AB}=0.96$.}
 \end{figure}

 \begin{figure}[htb]
 \centering \includegraphics[width=0.47\textwidth]{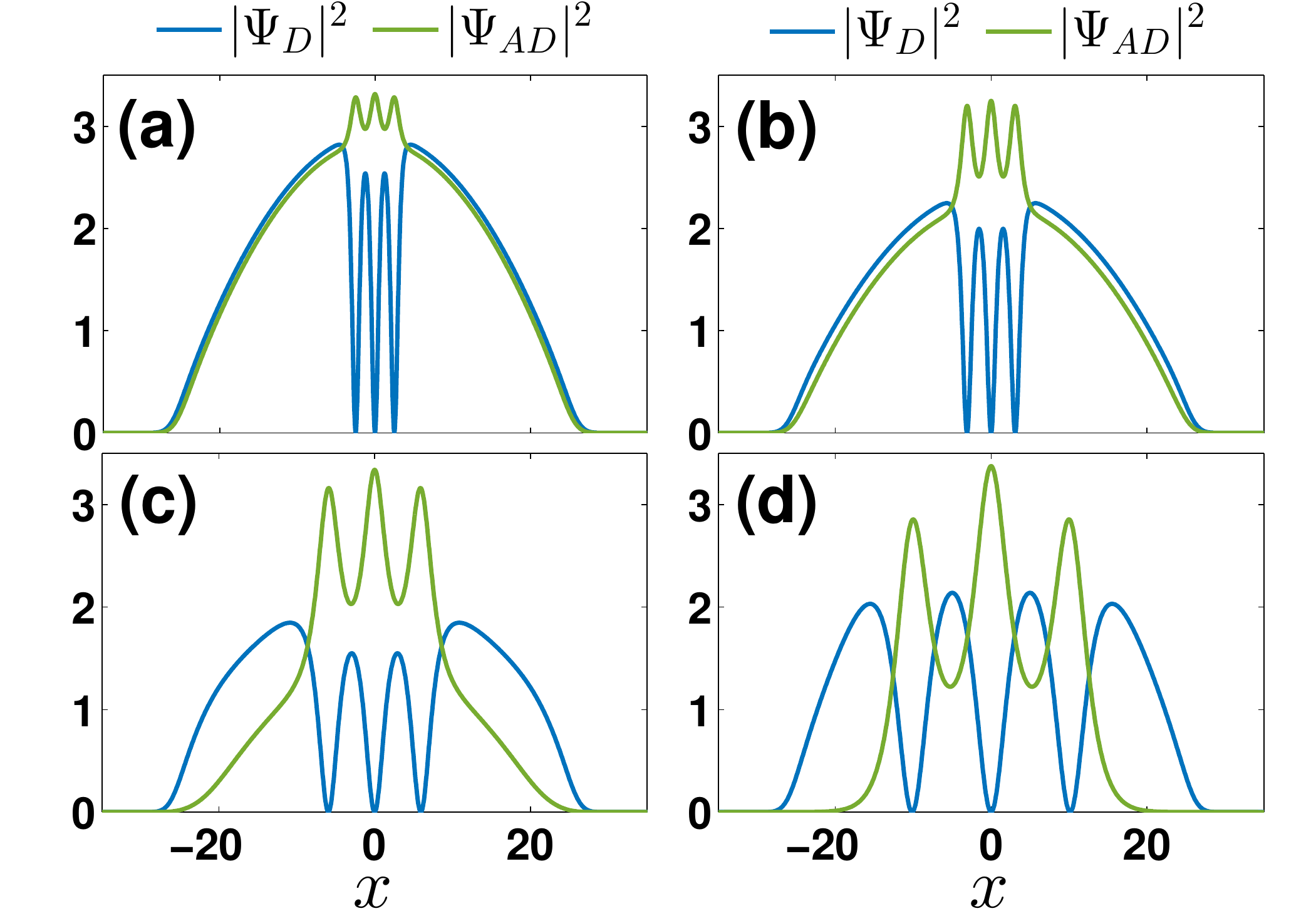}
 \caption{\label{Figure_3DADs_stat} Stationary states of
 three sorted DAD solitons for different interspecies interactions, namely (a) $g_{AB}=0.2$, (b) 
 $g_{AB}=0.5$, (c) $g_{AB}=0.9$ and (d) $g_{AB}=0.98$.
 The density of both the dark, $|\Psi_D|^2$, and the antidark, $|\Psi_{AD}|^2$, component is 
 shown (see legend).
 The presented states remain stable for long evolution times up to $t=10^4$. 
 The remaining parameters of the system are $\omega_1=0.1$, $g_{AA}=1.004$, $g_{BB}=0.9898$, $
 \mu_A=3.5$ and $\mu_B=3.4$.}
 \end{figure}

 \begin{figure*}[htb]
 \centering \includegraphics[width=0.9\textwidth]{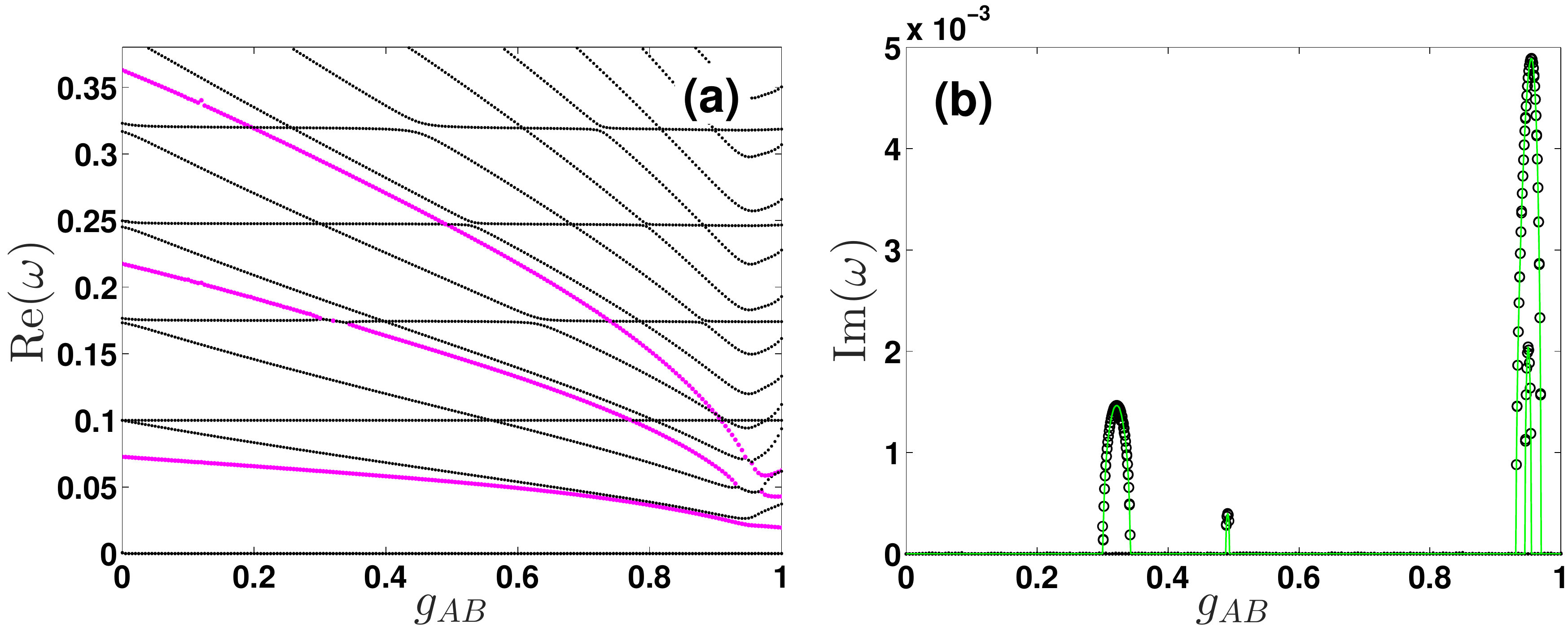}
 \caption{BdG spectrum of a stationary state of three sorted DAD solitons with 
respect to the interspecies interaction strength $g_{AB}$. (a) Real part, ${\rm Re}(\omega)$, 
of the underlying eigenfrequencies for varying $g_{AB}$. 
The points lying at ${\rm Re}(\omega)=0$ are quadruple degenerate due to particle number conservation in each component. 
(b) Imaginary part, ${\rm Im}(\omega)$, of the eigenfrequencies.
The corresponding instability windows of the three-DAD configuration
are $0.301 \leq g_{AB} \leq 0.342$, $0.489 \leq g_{AB} \leq 0.493$
and $0.932 \leq g_{AB} \leq 0.969$.
Solid lines in panel (b) provide a guide to the eye for each of the individual bifurcation loops  
emerging in the imaginary part of the eigenfrequencies.
The remaining system parameters used correspond to $\omega_1=0.1$, $\mu_A=3.5$, $\mu_B=3.4$, 
$g_{AA}=1.004$ and $g_{BB}=0.9898$.
\label{Figure_3DADs_BdG_g12} 
}
 \end{figure*}

 \begin{figure}[htb]
 \centering \includegraphics[width=0.47\textwidth]{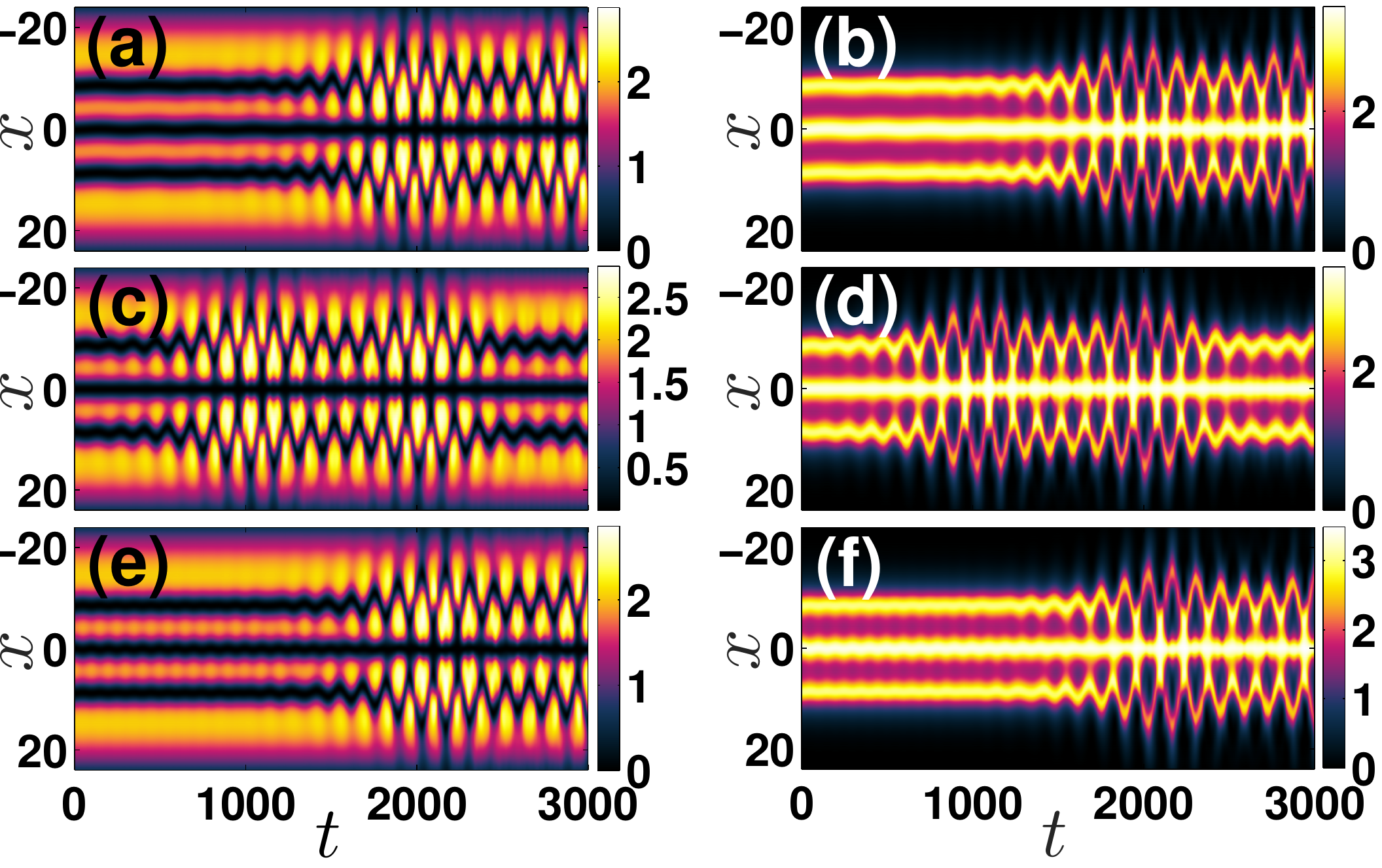}
 \caption{\label{Figure_3DADs_096} (a)-(f) {Dynamical evolution of three sorted DAD solitons.}
 Panels (a), (c), (e) [(b), (d), (f)] present the evolution of the dark [antidark] soliton
 component upon adding respectively the eigenvectors corresponding to the first [(a), (b)], the
 second [(c), (d)] and the third [(e), (f)]
 anomalous mode identified in the BdG spectrum of Fig. \ref{Figure_3DADs_BdG_g12}.
 The remaining system parameters correspond to $\mu_A=3.5$, $\mu_B=3.4$, $g_{AA}=1.004$, $g_{BB}=0.9898$ and
 $g_{AB}=0.96$.}
 \end{figure}

For all of our numerical investigations a fixed-point
numerical iteration scheme is employed~\cite{NewtonKrylov} in order to obtain bound states consisting of multiple DADs.
To simulate the dynamical evolution of the DAD arrays governed by Eqs.~(\ref{eqn::GPE_A})-(\ref{eqn::GPE_B}),
a fourth-order Runge-Kutta integrator is employed and a second-order finite difference
method is used for the spatial derivatives.
The spatial and temporal discretization spacings are chosen as
$dx = 0.05$ and $dt = 0.001$
respectively.
Moreover our numerical computations are restricted to a finite region
by employing hard-wall boundary conditions.
The latter are chosen wide enough in order to avoid finite size effects.
Particularly, in the dimensionless units adopted herein, the hard-walls are located at $x_{\pm}=\pm 80$ and
we do not observe any appreciable density for $\abs{x} > 30$.

\section{Sorted dark-antidark solitons: Stability analysis and dynamics}\label{sorted}

In the following we will explore the stability of bound states consisting of an arbitrary number of dark solitons in the first
component of the mixture and corresponding antidark ones building upon the second component of the binary system of
Eqs.~(\ref{eqn::GPE_A})-(\ref{eqn::GPE_B}).
These configurations with all bright (or antidark) solitonic features in one component and all dark solitonic features in the other component will
be referred to in the following as sorted DAD arrays.
We remark that bound states composed of alternating dark and antidark entities within the same component
cannot exist as stationary configurations, see also Sec. \ref{alternating}.
While such alternating states can (and do) emerge through the
experimental procedure used in Sec.~\ref{sec:experiment}, nevertheless they
correspond to dynamically evolving (rather than stationary) states of the system.

In order to obtain the sorted stationary states, a $\tanh$-shaped profile is used as an initial ansatz
for the wavefunction with $n_S$ dark solitons that reads
\begin{eqnarray}
\Psi_D= A(x) \prod^{n_S}_{j=1}  \tanh\left[D\left(x-x_{0_j} \right) \right]. \label{dark_soliton_state}
\end{eqnarray}
In the above expression $A(x)=\left(1/\sqrt{g_{AA}}\right) \sqrt{\mu_A-V(x)}$ is the
common Thomas-Fermi background into which the dark solitons are embedded.
Additionally, $D$ and $x_{0_j}$ refer to the common inverse width and the center of
the $j$-th dark soliton respectively.
The ansatz employed for the initial guess wavefunction of the corresponding antidark states is
\begin{eqnarray}
\Psi_{AD}=  \sum^{n_S}_{j=1} B(x)+C\sech\left[D\left(x-x_{0_j} \right) \right],\label{DAD_state}
\end{eqnarray}
where the relevant background here is given by $B(x)=\left(1/\sqrt{g_{BB}}\right) \sqrt{\mu_B-
V(x)}$, and $C$ denotes the amplitude of the density peak (on top of the
background).
Recall that DAD states consist of a density hump modeled here by a bright soliton as in the second 
term of Eq.~(\ref{DAD_state}) on top of the
BEC background \cite{Danaila,Kevrekidis_families,Katsimiga_MB_DBs,Mistakidis_cor_effects,Kiehn}.
Utilizing the above ans{\"a}tze, stationary states consisting of an arbitrary number of sorted DAD 
solitons symmetrically placed around the origin ($x=0$) have been identified.

\subsection{Two sorted DADs}

Prototypical examples for states with two sorted DADs are shown in Fig.~\ref{Figure_2DADs_stat}.
Here, we see how the relevant state changes upon variation of the intercomponent interaction coefficient $g_{AB}$.
The two-dark solitons that are well-known to form a bound state in single-component
BECs~\cite{siambook,Frantzeskakis_2010}
now produce an attractive potential (due to the repulsive nature
of the interaction and the absence of atoms in the dark solitons) for
the second component.
Thus, in this potential well, atoms are trapped on
top of the background state of the second component, generating
antidark states.
As $g_{AB}$ is gradually increased these states progress towards the
immiscible limit and eventually in the vicinity of the latter
threshold ``isolate'' themselves into bright-soliton-like droplets.
Indeed, as the immiscibility threshold is crossed, the second component
only prefers to localize itself in these bright structures, suggesting
a morphing of the multi-DAD states into multi-DB ones.

To assess the stability of the aforementioned stationary DAD states, we perform
a Bogoliubov-de Gennes (BdG) analysis, linearizing around the
equilibrium as follows:
\begin{eqnarray}
  \Psi_D &=& \Psi_{D}^{(eq)} + \left(a(x) e^{-i \omega t} + b^{\star}(x) e^{i \omega^{\star} t}
  \right),
  \label{bdg1}
  \\
 \Psi_{AD} &=& \Psi_{AD}^{(eq)} + \left(c(x) e^{-i \omega t} + d^{\star}(x) e^{i \omega^{\star} t}
  \right).
  \label{bdg2}
\end{eqnarray}
The resulting linearization system for the eigenfrequencies $\omega$
(or equivalently eigenvalues $\lambda=i \omega$) and
eigenfunctions $(a,b,c,d)^T$ is solved numerically. If modes with purely
real eigenvalues (genuinely imaginary eigenfrequencies) or complex
eigenvalues (eigenfrequencies) are identified, these are tantamount
to the existence of an instability~\cite{Kevre}.
Indeed, upon our systematic variation of $g_{AB}$ as discussed above (which, for suitable atomic species, is experimentally realizable via the use of Feshbach resonances),
we identify such instabilities.
While the two-DAD state is dynamically stable for a wide range of parametric values, there exist narrow intervals
of $g_{AB}$ (mentioned in the caption of Fig.~\ref{Figure_2DADs_BdG_g12}) for which the solution is predicted to be unstable.

This suggests the relevance of a further effort in order to
identify the modes responsible for the existence of
the instability. We note that in addition to four modes in the spectrum
at $\lambda=\omega=0$ due to symmetries, namely the conservations of
the particle numbers in each component of the mixture, there are additional
modes of interest that are referred to as anomalous or negative energy
ones~\cite{Skryabin}. These are modes highlighting the excited nature
of
the state under consideration (i.e., for the ground state there are no
such modes).
These eigenstates are quantified via the so-called
negative energy or negative Krein signature~\cite{Skryabin}.
The mode energy (or Krein signature) is defined as
\begin{eqnarray}
  K= \omega_1 \int  \Big(|a|^2 - |b|^2 + |c|^2 - |d|^2\Big) dx,
  \label{krein}
\end{eqnarray}
in a multi-component system like the one considered herein.
An example of this sort is shown in the BdG spectrum presented in the left panel 
of Fig.~\ref{Figure_2DADs_BdG_g12} for the two sorted DAD configuration where the
negative energy modes are denoted by magenta dots.

The existence and parametric variation of such modes is of particular relevance since their
collision with opposite (positive) Krein signature modes gives rise to
stability changes in the form of oscillatory instabilities or
Hamiltonian-Hopf bifurcations~\cite{Skryabin,siambook}. 
Indeed what is happening here, along the axis of real
eigenfrequencies, is that modes with $K>0$ and $K<0$ collide in pairs and give
rise to complex eigenfrequency quartets, as we will observe below.
These resulting complex eigenfrequencies
(all four of them) have $K=0$ until they ``complete'' an oscillatory
instability ``bubble''. 
Then, these eigenmodes ``land back'' on the real eigenfrequency axis
and retrieve their respective $K<0$ or $K>0$ traits.
The resulting instability is the one 
that we will trace later on in the dynamics as well. 
Our calculations show that there are {\it two} such modes in the system consisting of two sorted 
DAD solitons.
Additionally, {\it three} and {\it six} such modes appear when considering
respectively the three- and the six-DAD soliton configuration; see,
e.g., Figs.~\ref{Figure_3DADs_BdG_g12} and \ref{Figure_BdG_6DADs} below.
This is in agreement with the case of dark solitons in single-component BECs, where an $N$-soliton 
state has been shown to possess $N$ negative energy modes~\cite{Kapitula}. 
Specifically, for all the distinct sorted DAD configurations investigated herein 
there exist intervals where the above-identified
anomalous modes collide with $K>0$ modes pertaining to the collective
excitations of the background condensate. 
For instance, such a collisional interval occurs for $g_{AB} \in [0.946, 0.967]$ for the two-DAD 
configuration [Fig.~\ref{Figure_2DADs_BdG_g12}(a)] while for the three- and six-DAD states such 
intervals appear e.g. for $g_{AB} \in [0.301, 0.342]$ [Fig.~\ref{Figure_3DADs_BdG_g12}(a)] and for 
$g_{AB} \in [0.19, 0.25]$ respectively [Fig.~\ref{Figure_BdG_6DADs}(a)].
Notice also that within the aforementioned parameter intervals, 
whenever such collisions take place a bifurcation occurs in the 
corresponding imaginary part presented in Figs.~\ref{Figure_2DADs_BdG_g12}(b), 
\ref{Figure_3DADs_BdG_g12}(b) and \ref{Figure_BdG_6DADs}(b) respectively
signaling the presence of an instability of the relevant DAD
configuration in each case.

We now explore the direct numerical evolution of the two-DAD solitonic state for different parametric values.
In Fig.~\ref{Figure_2DADs_036}, we offer an example for $g_{AB}=0.36$ for which the relevant configuration is stable.
We examine the dynamical outcome of the coherent structure when adding the first (top panels)
and the second (bottom panels) anomalous mode to the wave.
Close inspection indicates that the former lower frequency mode leads to an in-phase oscillation of the two
DAD structures, while the latter higher frequency mode leads to an out-of-phase one.
Nevertheless, both configurations turn out to be stable, in agreement with our stability analysis observations in
Fig.~\ref{Figure_2DADs_BdG_g12}.
On the contrary, in Fig.~\ref{Figure_2DADs_096}, the higher frequency anomalous mode associated with out-of-phase
DAD solitary wave oscillations is unstable due to a resonant collision with one of the positive Krein signature modes.
Consequently (and irrespectively of whether we excite chiefly the first (top panels)
or the second (bottom panels) anomalous mode), eventually an instability ensues.
Naturally, in the bottom panels where the responsible mode for the instability has been excited,
the relevant phenomenology arises earlier.
Nevertheless, it arises in both cases and leads to a resonant growth of the out-of-phase
DAD waves' oscillation amplitude before leading to a saturation and subsequently to a recurrence effect
(see especially the bottom panels of Fig.~\ref{Figure_2DADs_096}).
{It is important to comment at this point that for larger $g_{AB}$ values also the lowest lying anomalous mode
destabilizes, resulting in the appearance of the inner loop present in Fig.~\ref{Figure_2DADs_BdG_g12}(b).
The associated instability window here occurs for $0.949\leq g_{AB} \leq 0.959$.
Notice however, that the imaginary eigenfrequencies corresponding
to this bifurcation are suppressed (i.e., of much smaller growth rate)
when compared to the
predominant ones of the outer bifurcation.
This suggests that even though the lowest-lying mode destabilizes, it
is not the one (principally) responsible
for the observed instability of the two-DAD configuration.
This
is indeed confirmed by inspecting once more the dynamical evolution presented in the top panels of
Fig.~\ref{Figure_2DADs_096}.}

\subsection{Three sorted DADs}

We now turn to a similar set of results for the configurations with three sorted DADs. In Fig.~\ref{Figure_3DADs_stat}, we
observe how the three solitary wave state progressively transforms
itself (in its stationary form) as $g_{AB}$ is increased. The
prevailing
picture at low $g_{AB}$ once again involves the dark solitary
waves forming wells where the antidark states of the second
component are trapped. Gradually, as $g_{AB}$ is increased,
more of the second component density gets encompassed in
the antidark states which eventually (as the immiscibility threshold
is approached) become essentially droplet like and will separate
into DB states for sufficiently large $g_{AB}$. As may be
expected, the BdG spectrum features now three anomalous modes
as illustrated in Fig.~\ref{Figure_3DADs_BdG_g12}. These, in turn,
yield their own potential resonant intervals as discussed in the
caption of the associated figure. The lowest of these modes
is again an in-phase one, the second mode involves a quiescent central DAD, while the outer ones oscillate out-of-phase;
finally, the third mode involves an in-phase oscillation of the outer waves and an out-of-phase one of the middle wave.
{The first mode is never resonant in the parametric regime
  considered.  The second mode causes the first instability window and
  is principally responsible for the third window (where it is associated
  with the highest growth rate), while the third mode causes the second
  instability window and also leads to a partial destabilization
  within the third window.
Focusing our attention on the predominant third bifurcation interval
  (associated with
  the maximal instability growth rate) present for the three DAD configuration
[Fig.~\ref{Figure_3DADs_BdG_g12}(b)],
we next examine the unstable dynamics associated with it.
Notice that as in the investigation of configurations with two DADs, here an inner loop appears in the respective BdG
spectrum [Fig.~\ref{Figure_3DADs_BdG_g12}(b)],
suggesting in this case the destabilization of not only the second
but also the third anomalous mode for values of $g_{AB}$
lying in the interval $0.946 \leq g_{AB} \leq 0.954$.
However, in Fig.~\ref{Figure_3DADs_096}, we can observe that independent of which of the modes we add to the initial stationary configuration with three DADs, the resonant second mode will
eventually be excited giving rise to the growth of the associated
configuration involving the out-of-phase motion of the outer
DADs while the middle one remains quiescent. The saturation
of the relevant growing oscillation and the recurrence of the
phenomenon are clearly observed especially in the middle
panel of the figure where the unstable mode was added initially
to the stationary state and hence gave rise to the associated
instability earlier.}

 \begin{figure}[htb]
 \centering \includegraphics[width=0.47\textwidth]{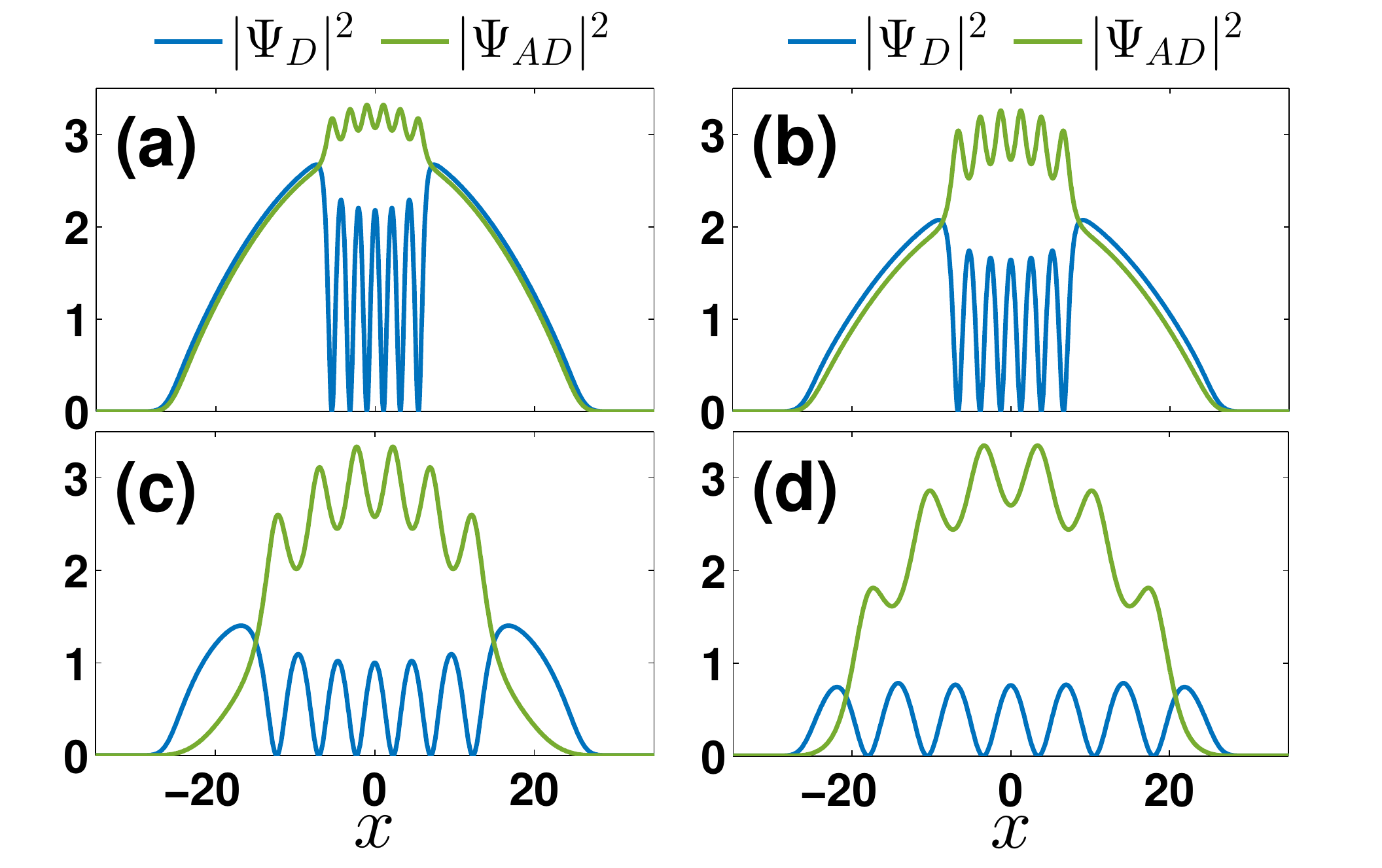}
 \caption{\label{Figure_6DADs_stat} Stationary states consisting of
 six sorted DAD solitons for different interspecies interactions, 
 namely (a) $g_{AB}=0.2$, (b) $g_{AB}=0.5$, (c) $g_{AB}=0.9$ and (d) $g_{AB}=0.98$.
 The densities $|\Psi_D|^2$ of the dark and $|\Psi_{AD}|^2$ of the antidark 
 components are shown (see legend).
The remaining system parameters are $\omega_1=0.1$, $g_{AA}=1.004$, $g_{BB}=0.9898$, $\mu_A=3.5$ 
and $\mu_B=3.4$.}
 \end{figure}

\begin{figure*}[htb]
 \centering \includegraphics[width=0.9\textwidth]{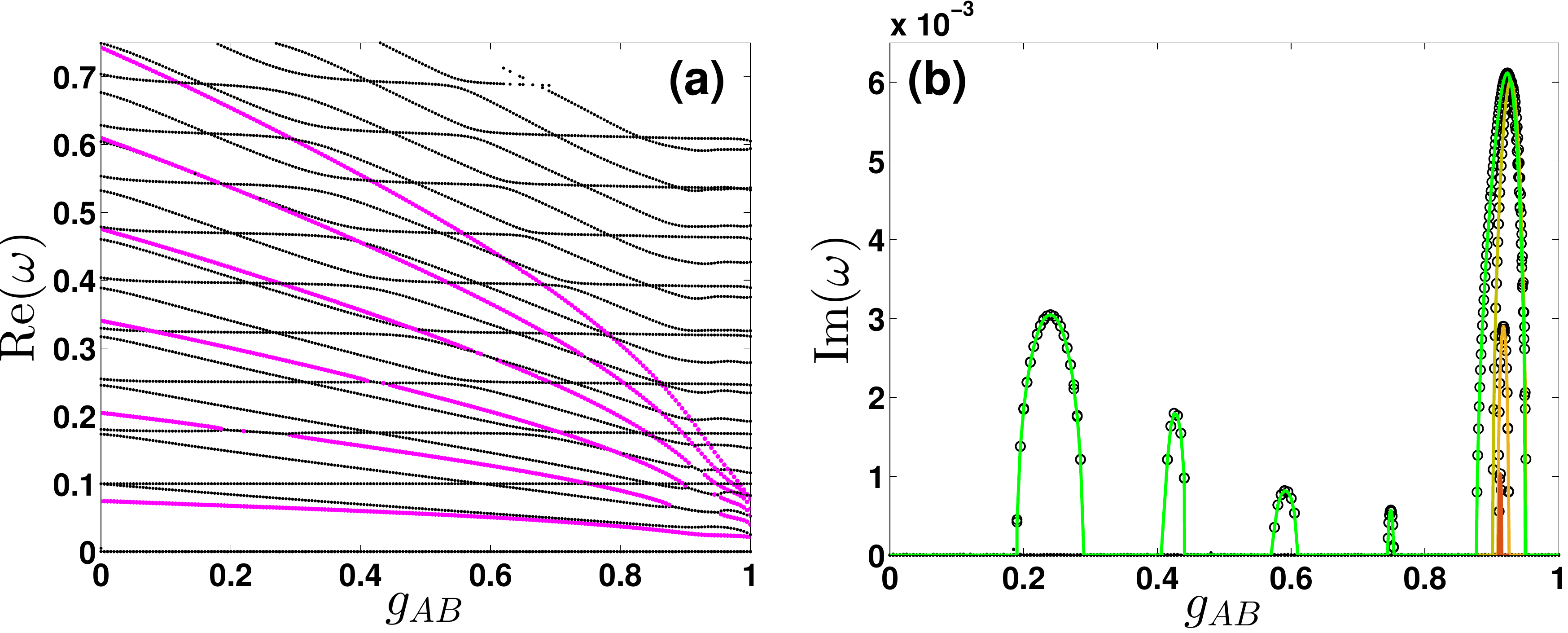}
 \caption{BdG spectrum of a stationary state consisting of six sorted DAD
 solitons. The trajectories of six anomalous modes
 can be inferred in (a) being denoted by magenta dots. 
 All points residing at ${\rm Re}(\omega)=0$ exhibit a quadruple degeneracy stemming from the particle number conservation 
of each component. 
Additionally, a cascade of bifurcations occurs, as $g_{AB}$ is varied, 
in this BdG spectrum as it is evident by the finite imaginary eigenfrequencies shown in (b). 
The associated instability intervals are: $0.19 \leq g_{AB} \leq 0.285$, 
$0.415 \leq g_{AB} \leq 0.44$, $0.575 \leq g_{AB} \leq 0.605$, $0.745 \leq g_{AB} \leq 0.752$, 
and $0.878 \leq g_{AB} \leq 0.951$. Solid lines shown in (b) provide a guide 
to the eye for the individual bifurcation loops appearing in the imaginary part of the eigenfrequencies.
Other parameters used are $\mu_A=3.5$, $\mu_B=3.4$, $g_{AA}=1.004$, and $g_{BB}=0.9898$.
 \label{Figure_BdG_6DADs}}
\end{figure*}

\subsection{Six sorted DADs}

In order to generalize our findings to even larger DAD soliton arrays we now consider the
stability properties of a stationary state consisting of six sorted DAD
solitons.
The gradual transformation of the obtained stationary states as $g_{AB}$
increases is illustrated in panels (a)-(d) of
Fig.~\ref{Figure_6DADs_stat}.
One can see in the figure how the blobs corresponding to the location
of the
individual DADs separate as $g_{AB}$ is increased.
To address the stability of these bound states we employ the same
diagnostics as in the previous two cases.
Specifically for the numerical findings to be presented below we fix $\mu_A=3.5$, $\mu_B=3.4$
and we vary $g_{AB}$ within the interval $[0, 1]$.
The relevant BdG spectrum is illustrated in Figs.~\ref{Figure_BdG_6DADs}(a) and (b).
Once again, the magenta dots are used to denote the anomalous
(negative Krein signature) modes in the figure.
Importantly, a cascade of Hamiltonian-Hopf bifurcations can be observed
in this BdG spectrum as is corroborated  by the finite imaginary eigenfrequencies
(or instability growth rates), ${\rm Im} (\omega)$, depicted in
Fig.~\ref{Figure_BdG_6DADs}(b).
The associated instability intervals, five in this case, are indicated in the figure caption.

The corresponding modes are simply the normal modes of vibration of the six solitary waves 
considered as particles.
{In that vein, the lowest out of the six anomalous modes present for the state with six DADs 
is an in-phase one. Accordingly, the highest, i.e. the sixth, anomalous mode involves 
an out-of-phase motion of adjacent coherent 
structures. Finally all the intermediate modes, i.e. from the second till the fifth one, entail 
relevant mixed phase oscillations.}
 \begin{figure}[htb]
 \centering \includegraphics[width=0.47\textwidth]{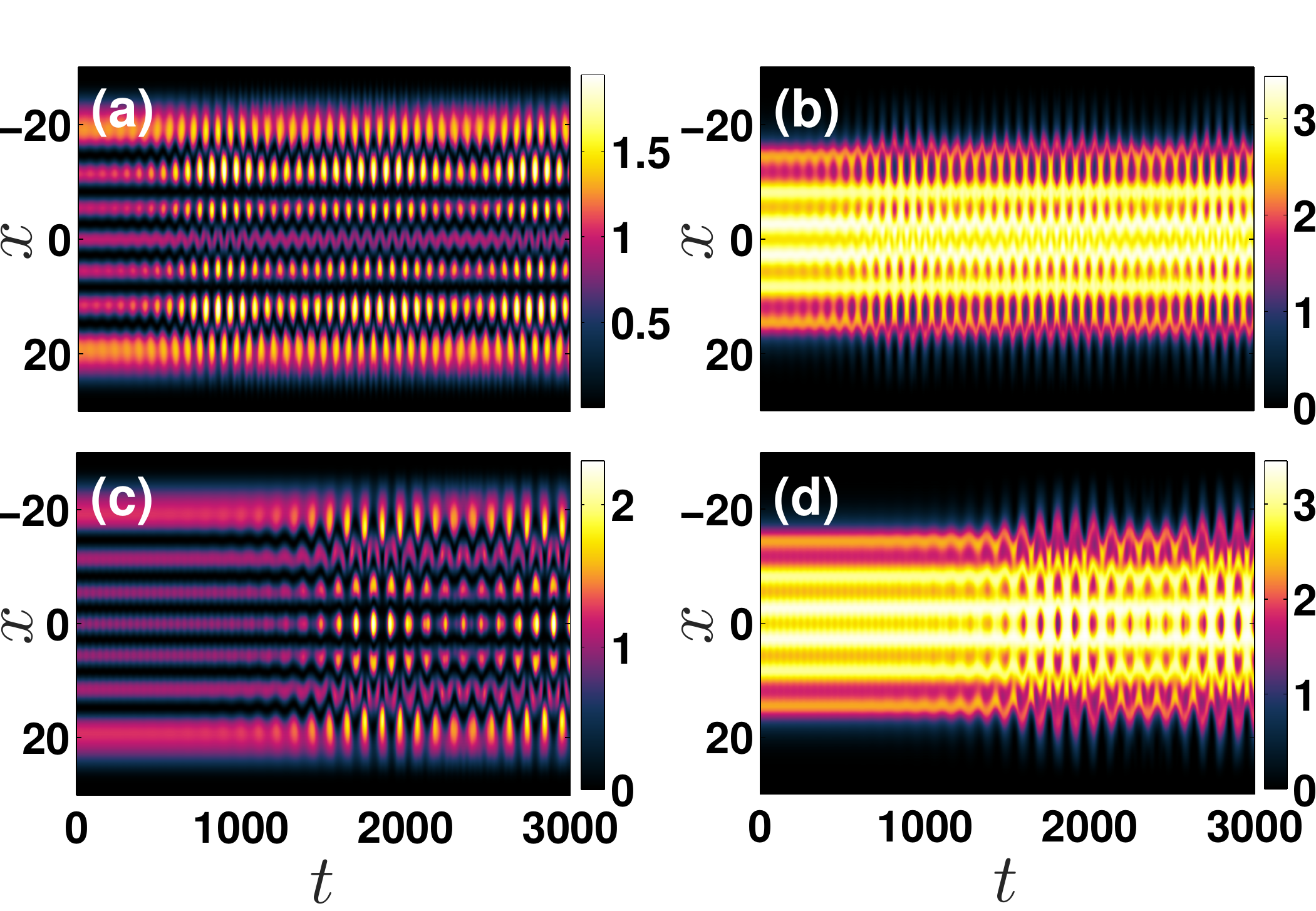}
 \caption{\label{Figure_6DADs_094} (a)-(d) Evolution of a stationary
   state of six sorted DAD solitary waves for $\mu_A=3.5$.
  The time-evolution of the dark [antidark] soliton component is presented 
  in panels (a), (c) [(b), (d)] upon adding the eigenvectors of the third [(a), (b)], 
  and the sixth [(c), (d)] anomalous mode identified in the BdG spectrum 
  of Fig. \ref{Figure_BdG_6DADs}.
  The remaining system parameters are $\mu_B=3.4$, $g_{AA}=1.004$, $g_{BB}=0.9898$ 
  and $g_{AB}=0.94$.}
 \end{figure}
Turning to the fifth interval, exhibiting the largest instability
growth rate, as illustrated in Fig.~\ref{Figure_BdG_6DADs}(b)
we next inspect the unstable dynamics of the six-DAD configuration for $g_{AB}=0.94$ 
(Fig.~\ref{Figure_6DADs_094}).
Before delving into the details of the associated dynamics,
it is important to stress at this point that
as the immiscibility threshold is approached, four out of the six
anomalous modes destabilize. This can be observed by close inspection of the spectrum, revealing 
four loops in the imaginary part of the BdG
spectrum in Fig.~\ref{Figure_BdG_6DADs}(b).
Two possible manifestations of the instability, using perturbations
along the third and sixth eigenmode are shown in Fig.~\ref{Figure_6DADs_094}.
Specifically, upon adding the third of the aforementioned modes to the initially stationary six-DAD state (top panels in Fig.~\ref{Figure_6DADs_094})
we observe that the instability manifests itself from the very early
stages of the dynamics leading to the oscillatory motion of the six-DAD configuration,
involving the in-phase oscillation of the central and the outermost
DAD waves. The intermediate pair of DADs remains approximately quiescent
during this oscillatory dynamical evolution.
\begin{figure}[htb]
\centering \includegraphics[width=0.5\textwidth]{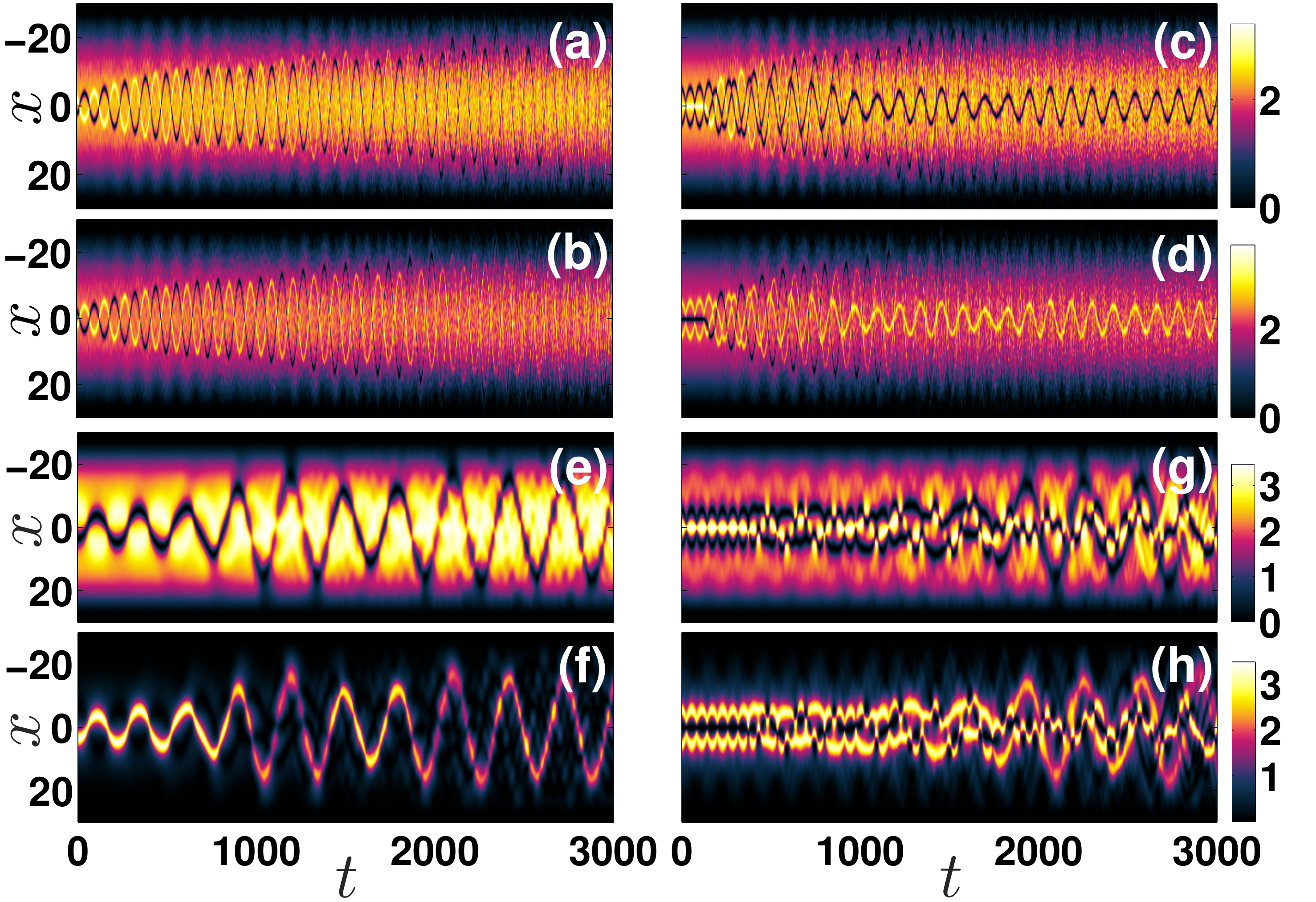}
\caption{Spatiotemporal evolution of the densities of both components
showcasing the oscillations and interactions of alternating DAD solitons for (a)-(d) $g_{AB}=0.6$
and (e)-(h) $g_{AB}=0.96$ respectively.
Panels (a), (e) [(c), (g)], correspond to an initial seeding of a single dark [two darks] and a 
single antidark in the first component and (b), (f) [(d), (h)] their relevant mirror images in the 
second component.
Other parameters used are $\mu_A=3.5$, $\mu_B=3.4$, $g_{AA}=1.004$, and $g_{BB}=0.9898$.
\label{Figure_Alter}}
 \end{figure}
However,  for all the remaining cases it is found that independently of which mode we add to the 
initially stationary six DAD state,
namely either the first, the second, the fourth, the fifth or the sixth mode
(bottom panels in Fig.~\ref{Figure_6DADs_094}), the resonant second mode
will eventually be excited giving rise to a growth of the corresponding configuration
that entails the out-of-phase vibration of the central DAD waves and the in-phase oscillation
between the second and the outermost DAD pairs.

\section{Alternating Dark-antidark solitons}\label{alternating}

Having discussed the static properties of arrays consisting of
multiple sorted DAD
solitons, motivated by the experimental observations
we next turn our attention to the dynamics of solitonic states consisting of dark and antidark waves in each of the components of the mixture in an alternating fashion.
Despite our extensive efforts to identify stationary states consisting of alternating dark and antidark solitons
(i.e. DAD configurations where the dark features, and in a complementary way the antidark features, alternate between the two pseudo-spin components) our numerical findings suggest that such a state cannot be stationary.
We have dynamically constructed such alternating states and monitored their evolution and interactions.
{Examples illustrating the spatiotemporal evolution of solitonic entities composed,
for instance, of one dark followed by an antidark soliton in the first component and vice versa in the second
component are presented in the first column of panels in Fig.~\ref{Figure_Alter} for $g_{AB}=0.6$ (top doublet of
panels) and  $g_{AB}=0.96$ (bottom doublet of panels) respectively.
Notice that regardless of the value of the intercomponent coupling strength, the alternating
entities perform oscillations of growing amplitude during the initial stages of the dynamics.
However, for smaller values of $g_{AB}$ and for times around $t\approx 2\times 10^3$ the alternating states
are lost within the significantly excited background and only a single
DAD soliton appears to survive
at later times. On the other hand, a distinct evolution possibility
arises as $g_{AB}$ increases.
For instance, for $g_{AB}=0.96$ and focusing on these later times, beating dark-dark solitons
develop in the two components and propagate within the BEC medium for
(dimensionless) times up to
$t\sim 3\times 10^3$ that we have considered~\cite{azu}.}

{A similar outcome (i.e., involving the unstable dynamics) arises
upon increasing the number of alternating waves in each component.
Specifically, and as depicted in the second column of Fig.~\ref{Figure_Alter},
when two dark (antidark) and a central antidark (dark) solitary waves
are initialized in the first (second)
hyperfine state, a nearly bound state formation occurs in the relevant dynamics.
Here, the central wave in each component oscillates in its
amplitude persistently at the center of the configuration while the outer two waves
perform out-of-phase oscillations. However, at later times,
i.e. around $t\approx 200$ for $g_{AB}=0.6$ and around $t\approx 500$
for $g_{AB}=0.96$, the alternating structures are set in motion resulting
in turn into a single sorted DAD soliton oscillating around the trap
center for all times in the former case, and oscillating and interacting beating dark-dark entities for the latter scenario.
}

{It is also worth commenting at this point that since the alternating states are not stationary ones
emission of radiation from the relevant pattern
takes place right after the initial (at $t=0$) seeding,
resulting in an excited BEC background in all cases studied here.
Finally, we note in passing that one can systematically study the dynamical evolution of the system
when considering an arbitrarily large number of alternating dark and antidark waves.
Here, more complex interactions between the
ensuing waves take place, including their deformation into beating dark-dark solitons (results not shown here).}

\section{Conclusions}\label{conclusions}

In the present work we have been motivated by experimental
realizations of two-component BECs in an elongated trap to consider
solitary wave structures involving dark solitonic states in one of
the components and corresponding antidark ones in the other. The
combination of a spatially dependent Zeeman shift with a
uniform fixed frequency microwave drive was demonstrated as a viable experimental pathway to the formation of such structures. It was possible
to create experimentally both cases where all the dark solitons were in
one component and the antidark ones in the other, as well as settings
where in each component the dark and antidark waves alternate.
We corroborated these experimental realizations with a theoretical
analysis based on a two-component Gross-Pitaevskii model in
a quasi-one-dimensional geometry. We saw that such a setup
enables the formulation of stationary configurations with all
the dark solitary waves in one component and the antidark
ones in the other component. In fact, we found such states
with two, three, six and in principle arbitrary ``lattices'' of dark waves
in one component and antidarks in the other.
On the other hand, this was not the case for the alternating
dark-antidark configurations. Such a state could only be traced
as a dynamical one and never as a stationary one.
For the multiple sorted DAD wave case,
we found that the waves were generally dynamically stable,
although the presence of $N$ anomalous modes (in the
states with $N$ DAD waves) could potentially lead to
windows of oscillatory instabilities. The latter were observed to give rise
to resonant growth of the oscillations involving the DAD waves,
but eventually a saturation thereof and a potential recurrence subsequently
of the associated dynamics.

Naturally, there are many possible extensions of this direction
of studies. From an analytical standpoint, it is relevant to
extend the type of understanding that exists for
the interactions of dark~\cite{Frantzeskakis_2010} and
even dark-bright~\cite{siambook,stockhofe} solitary
waves to the realm of dark-antidark structures.
This will provide a guideline for understanding the
formation of equilibria (when all the darks are in the same
component) or the absence thereof (when the adjacent
darks are in alternating components).
Another direction that would be of substantial
interest would be to extend relevant structures to
the realm of spinor condensates with three spin states where it is possible
to envision different types of extensions, e.g., ones
where two components are dark and one antidark,
as well as ones where only one component is
dark and two are antidark~\cite{schmied}.
Finally, it is also natural to extend considerations
to higher dimensions and seek lattices of multiple
vortex-antidark states either with the vortices bearing
the same or alternating topological
charges. Understanding in a quantitative fashion
the interaction of two such states or the formation of lattices of more such states is also
a topic of ongoing interest~\cite{kasamatsu}.
Studies along these directions are presently underway and
will be reported in future publications.

\section*{Acknowledgements}
P.G.K. is grateful to the Leverhulme Trust and to the
Alexander von Humboldt Foundation for
support and to the Mathematical Institute of the University of Oxford
for its hospitality. This
material is based upon work supported by the US National Science
Foundation under Grants PHY-1602994 and
DMS-1809074 (PGK)
G.C.K., S.I.M. and P.S. gratefully acknowledge financial support by the Deutsche
Forschungsgemeinschaft (DFG) in the framework of the SFB 925 ``Light induced dynamics and control
of correlated quantum systems".
S.I.M. gratefully acknowledges financial support in the framework of the Lenz-Ising Award of
the University of Hamburg. T.M.B, S.M.M., M.K.H.O., and P.E  acknowledge funding from the NSF under grant numbers PHY-1607495 and PHY-1912540.

{}

\end{document}